\documentstyle[aps,epsf,twocolumn,prb,amssymb]{revtex}

\def\probe{CuGa$_2$O$_4$}
\def\musr{$\mu$SR}

\begin{document}
\draft
\wideabs{

\title{Spin-glass state in \probe}
\author{G.A. Petrakovskii, K.S. Aleksandrov, L.N. Bezmaternikh, S.S. Aplesnin}
\address{Institute of Physics, Academy of Sciences, 
Siberian Branch, 660036 Krasnoyarsk, Russia}
\author{B. Roessli, F. Semadeni}
\address{Laboratory for Neutron Scattering, 
Paul Scherrer Institute and ETH Zurich, CH-5232 Villigen PSI, Switzerland}
\author{A. Amato, C. Baines}
\address{Laboratory for Muon-Spin Spectroscopy, 
Paul Scherrer Institute, CH-5232 Villigen PSI, Switzerland}
\author{J. Bartolom\'e, M. Evangelisti*}
\address{Instituto de Ciencia de Materiales de Arag\'on, CSIC-Universidad de Zaragoza, Ciudad Universitaria,
50009 Zaragoza, Spain}
\date{\today}
\maketitle
\begin{abstract}
\indent
Magnetic susceptibility,  magnetization, specific heat and 
positive muon spin relaxation (\musr) measurements have been used to 
characterize the magnetic ground-state of the spinel compound $\rm CuGa_2O_4$. 
We observe a spin-glass transition of the S=1/2 $\rm Cu^{2+}$ spins
below $\rm T_f=2.5K$ characterized by a cusp in the susceptibility curve which is
suppressed when a magnetic field is applied.
We show that the magnetization of $\rm CuGa_2O_4$ depends on the magnetic history of the sample.
Well below $\rm T_f$, the muon signal resembles the dynamical Kubo-Toyabe
expression reflecting that the spin freezing process in $\rm CuGa_2O_4$ results in a 
Gaussian distribution of the magnetic moments. By means of Monte-Carlo simulations,
we obtain the relevant exchange integrals between the $\rm Cu^{2+}$ spins in
this compound.
\end{abstract}


\noindent
}
\section{Introduction}
\label{intro}
Although spin glasses have been extensively studied in the past years, there is 
still no consensus about the
ground-state and dynamics in these systems (for an introduction see e.g. 
K.H. Fisher and J.A. Hertz  [\onlinecite{Fisher}]). It is generally
accepted that both site-disorder and competition between the magnetic moments 
are necessary to produce a
low-temperature state where the spins are frozen along arbitrary directions
\cite{Fisher2}. 
Examples of such systems are
metallic spin-glasses where magnetic impurities are randomly diluted in a noble 
metal \cite{Souletie}. For
this particular class of materials competition between the magnetic moments is 
the result of the
Rudermann-Kittel-Kasuya-Yosida (RKKY) interaction \cite{RKKY} where 
ferromagnetic and antiferromagnetic
exchange interactions alternate as a function of distance between neighboring 
spins. The RKKY interaction
cannot be invoked for localized magnets and the spin-glass transition in these 
systems must be realized by
other mechanisms. Typical insulating spin-glasses of this kind are the alloys 
Eu$_x$Sr$_{1-x}$S. In the
$x$=0-limit, EuS is a well-known example of an isotropic 3-dimensional 
Heisenberg ferromagnet. The exchange
integrals have been determined by inelastic neutron scattering in this material 
with the result that
ferromagnetic nearest-neighbor exchange interaction competes 
with next-nearest antiferromagnetic coupling \cite{Bohn}. Diluting non-magnetic 
Sr for Eu ensures
bond-randomness and the conditions for obtaining a spin-glass state are 
fulfilled in a large range of
impurity concentrations \cite{Maletta}, in qualitative agreement with the 
molecular-field theory of Edwards
and Anderson \cite{EA}. De Seze pointed out that a spin-glass phase transition 
can occur in a geometrically
frustrated system with Ising spins and antiferromagnetic interactions only 
\cite{deSeze}. Following De Seze's work, Villain \cite{Villain} proposed  that spin glasses can be 
obtained in materials with geometric
frustration and Heisenberg-type exchange interactions like cubic spinels.  
These compounds have the chemical
formula AB$_2$O$_4$. The chemical structure of spinels consists of
both tetrahedral and octahedral sites. The  number of crystallographic sites
is larger than the number of A and B cations in the chemical formula, so that
the cations generally distribute randomly among the available atomic positions.
In particular, this random distribution of cations determines in a large extent
the microwave relaxation properties of the spinel compounds\cite{Sparks}.
 When both sublattices are occupied by magnetic ions the 
ground state is a ferrimagnet.
The B sublattice builds connected tetrahedra and antiferromagnetic interactions 
induce topological
frustrations 
\cite{Anderson} which can lead to a spin-glass state when non-magnetic 
impurities are introduced
\cite{Villain}. The dominant magnetic interaction in most of these materials is 
antiferromagnetic and connects
spins between the A and B sublattices while the A-A and B-B exchange 
interactions are 
comparatively small. However, intersublattice exchange constants can modify the 
magnetic phase diagram
originally calculated by Villain and for real systems the situation is usually 
complicated \cite{Pool}. Although
spin-glass transition has been found in diluted spinels \cite{Fiorani}, 
spin-glass transition in
pure cubic spinels is less common.

In this article, we report magnetic susceptibility, magnetization measurements
in fields up to 50 KOe,  specific heat and muon-spin relaxation ($\rm \mu SR$) 
measurements in the cubic spinel
CuGa$_2$O$_4$. The results show that $\rm CuGa_2O_4$ undergoes a paramagnetic to 
spin-glass phase-transition at $\rm T_f=2.5K$. By means of Monte-Carlo simulations, 
the relevant exchange interactions are obtained for $\rm CuGa_2O_4$.  We show
that the formation of a spin-glass ground-state in $\rm CuGa_2O_4$ is due to the 
Jahn-Teller character of the $\rm Cu^{2+}$ ions. 
Specifically, in a field of octahedral symmetry, the Jahn-Teller effect 
distorts the electronic $d$ levels of the $\rm Cu^{2+}$ which become 
split by the effect of the crystal field into a threefold degenerate level and a 
twofold degenerate one. In such compounds there is an interaction between the electronic 
system with the underlying lattice which very often leads to a structural phase transition. 
Typical compounds exhibiting cooperative Jahn-Teller distortion are found 
in \textit{e.g.} perovskites ($\rm KCuF_3$, $\rm LaMnO_3$), spinels 
($\rm CuFe_2O_4$, $\rm Mn_3O_4$), rutiles ($\rm CrF_2$, $\rm CuF_2$) 
or garnets ($\rm Ca_3Fe_2Ge3O_{12}$). The structural phase transition can be accompanied by 
orbital-ordering of the $d$-electrons which in turn influences 
the nature of the exchange interaction. The important role of the Jahn-Teller effect 
in forming the magnetic ground-state in the perovskite manganites which exhibit colossal 
magnetoresistance (\textit{e.g.} see [~\onlinecite{popovic}~] and references therein) 
and in cuprates (\textit {e.g.} [~\onlinecite{zhao,shengelaya}~] and references therein)
is currently a subject of intense investigation both theoretically and experimentally.    
In that respect, we note that the influence of the Jahn-Teller effect on the properties of the magnetic insulators is discussed in detail by Kugel and Khomsky\cite{kugel}.  
\section{Experimental details}
\subsection{Sample Preparation}
Single crystals of CuGa$_2$O$_4$ 
were grown by spontaneous
crystallization starting from a $\rm CuO-Ga_2O_3$ solution melt 
in $\rm PbO-0.64B_2O_3-0.5Na_2O$. After  slowly cooling the melt to room temperature,
single crystals of typical size $3\times 3\times 3$~mm$^3$ and
of octahedral shape were obtained. X-Ray diffraction analysis showed that 
the  CuGa$_2$O$_4$ crystals used for the present experiments are  cubic spinels 
with both copper and gallium ions randomly distributed in the A and B sublattices 
in agreement with  previous  diffraction investigation (see Ref. [\onlinecite{Gonzales}]). 
The chemical
structure of $\rm CuGa_2O_4$ is described by the space-group $Fd\bar{3}m$ with 
lattice constants a=8.39\AA\ at room temperature.

\subsection{Magnetic Measurements}
The magnetic susceptibility and magnetization measurements were performed with
a commercial MPMS Quantum Design SQUID magnetometer together with an AC-susceptibility option at 
ICMA, Spain. The amplitude of the AC-magnetic field was set to 4.5 Oe 
with the frequency
of the field being varied between 0.1 Hz and 990 Hz. 
The measurements were carried out in the temperature range 
1.7-300 K and in applied magnetic fields up to 50 kOe. 
Additional measurements of the magnetic susceptibility in 
the temperature range T=4.2K to 120K were performed at the Institute of Physics,
 Krasnoyarsk, using a home-built SQUID magnetometer.

\subsection{Specific Heat Measurements} 
The specific heat measurements  were performed with a commercial PPMS device
(Quantum Design) in the
temperature range 1.8~K$\le T \le 10$~K. We used a small single crystal of mass 
$\sim$4.15mg. The raw data
were corrected for the copper host and glue, which were measured separately. 
We did not attempt to subtract the phonon contribution, as it is
expected to be small at low temperatures.

\subsection{Muon-Spin Relaxation}
The \musr\ experiments were performed on the LTF spectrometer at the Paul-Scherrer 
Institute, Switzerland. The
data were recorded using the zero-field method which is very 
sensitive to determine both static and dynamic effects in spin-glasses 
\cite{Emmerich}. Additional
measurements were performed as a function of applied magnetic field. 
In that case, the sample was zero-field cooled.
The sample we used for the present experiment consists of about 50 pieces 
of the above 
described crystals which were glued on a silver plate. The sample was enclosed 
in a top-loading
$^3$He-$^4$He dilution cryostat and the measurements were carried out in the 
temperature range 650~mK $\le T
\le 10$~K.

\section{Magnetization, susceptibility and specific heat results}
Figure\ref{fig1} shows the result of the magnetic susceptibility measurements
with an AC-frequency of 19 Hz and an excitation amplitude of H=4.5 Oe.
 For temperatures
higher than T=20 K, the magnetic susceptibility is well reproduced by the
Curie-Weiss law. Upon lowering the temperature below T=20 K,
the magnetic susceptibility increases continuously. The real
part of the magnetic susceptibility shows a cusp at $\rm T_f\simeq 2.5K$ 
which is independent of the relative orientation of the magnetic
 field with respect to the crystal axes. 
The imaginary part of the magnetic susceptibility 
also exhibits a maximum at the same temperature. To understand the nature of
the maxima appearing in the susceptibility curves, 
the magnetization in $\rm CuGa_2O_4$
 was determined as a function of applied magnetic field and for different 
magnetic histories of the samples. As an example, Fig. \ref{fig2} shows
the magnetization curves obtained in $\rm CuGa_2O_4$ after  
zero-field and field-cooling, respectively. 
For the latter case, the sample was cooled in a
magnetic field of H=100 Oe applied along the [001] crystal axis. 
It is evident from the figure that for temperatures below $\rm T_f\simeq 2.5K$
the magnetization shows a temperature hysteresis, which is an usual characteristic
for the formation of a spin-glass state.
The results of the magnetic susceptibility measurements
taken for different magnetic fields are presented in Fig. \ref{fig3}
 which shows that magnetic fields larger than H=5 kOe suppress the cusp 
observed at  $\rm T_f$ in zero-magnetic field. Figure \ref{fig4} shows that
the increase of the magnetic moment as a function of magnetic field is far
from saturation at the maximum field of 5T. In   
Fig. \ref{fig5} the temperature dependence of the transition temperature $\rm T_f$
of the spin-glass transition is observed to increase as a function of 
increasing AC-frequency.     
The above experimental results all indicate that the $\rm Cu^{2+}$ magnetic moments
in $\rm CuGa_2O_4$ undergo a phase transition to a spin-glass ground state
below $\rm T_f\simeq 2.5 K$. This is also confirmed by the calorimetric 
measurements performed in zero-magnetic field for this compound.
A plot of the specific heat $C_p/T$ in $\rm CuGa_2O_4$ is
shown in Fig. \ref{Fig6}. The data do not show 
any indication of a phase transition to a 3-dimensional 
ferro- or antiferro-magnetic
ordered state. However, a broad maximum is observed
around $\rm T = 2.5$~K followed by a slow decay toward high temperatures. This particular 
behavior of the specific heat as a function of  
temperature is reminiscent
of a spin-glass transition \cite{Fogle}.
\section{Discussion of the bulk measurements}
A spin-glass state is characterized by an assembly of 
 magnetic moments which are frozen along random 
and arbitrary directions in space below a specific transition temperature $\rm T_f$. 
Because of the non-ergodicity of the system, the phenomenon is irreversible.
The macroscopic magnetization of a spin-glass system is equal to zero in the absence
of a magnetic field. On the other hand, cooling a spin-glass in an external magnetic
field transfers the system into a metastable state with a non-zero magnetization value.
For temperatures above the spin-freezing temperature $\rm T_f$, the magnetic moments
are in a paramagnetic state
 and consequently the temperature dependence of the magnetic susceptibility follows
the Curie-Weiss law
\begin{equation}
\chi (T)= {C \over {T-\theta}},
\end{equation}
where $ C=Ng^2\mu_B^2S(S+1)/ {3k_B}$ is the Curie constant and $\theta$
the paramagnetic Curie temperature. $N$ is the magnetic moment density, $g$
 the Land\'e
factor and $\mu_B$ the Bohr magneton. $S$ corresponds to the spin value of
 $\rm Cu^{2+}$ and $k_B$ is the Boltzmann constant. From the 
magnetic susceptibility measurements presented above, we obtain  for $\rm CuGa_2O_4$
 the values $1/C$=0.34 emu K/mol, $ \theta = -8$ and 
$\rm \mu_{eff}=g\sqrt{S(S+1)}\mu_B = 1.65 \mu_B$.
The magnetic susceptibility is related to the Edwards-Anderson (EA) parameter\cite{EA}
$\rm q=\lim_{t \rightarrow \infty} [\langle S_i(t)S_i(0)
\rangle]_{av.}$ through the relation
\begin{equation}
\chi(T)=C{{1-q(T)} \over {T-\theta (1-q(T))}}.
\end{equation}
According to the percolation theory of Kirkpatrick\cite{kirk},
the EA-parameter follows a power law
$\rm q(T) \propto (1-T/T_f)^\beta$ close to the spin-glass temperature $\rm T_f$ with
 $\beta$ equal to 0.39. However, near $\rm T_f$, we found the
value $\beta$ = 0.16 in $\rm CuGa_2O_4$. 
The frequency dependence of the spin freezing temperature $\rm T_f$ is
a characteristic feature of the spin-glass state. It has been experimentally
found that in spin-glasses 
 $\rm T_f$ decreases with increasing AC-frequency. A quantitative measure of
the frequency shift is obtained from 
$\rm (\Delta T_f/T_f)/\Delta log(\omega)$=0.067. This is very similar to
other insulating spin-glasses such as $\rm (EuSr)S$ or $\rm (FeMg)Cl_2$
[\onlinecite{Mydosh}] and one order of magnitude lower than in canonical 
metallic spin glasses, or one order of magnitude larger than in a
superparamagnet.  
The magnetic field and temperature dependence of the magnetization M(H,T) for a spin-glass
system with Heisenberg spins of dimension $m$ 
has been calculated by Toulouse and Gabau\cite{toulouse}.
Within the Sherrigton-Kirkpatrick (SK) model and
for temperatures below $\rm T_f$, the expression for the magnetization is 
accordingly given by
\begin{equation}
M/H=1-[4/(m+2)]^{1/3}3h^{4/3}/4
\label{mh}
\end{equation}   
where $h=g\mu_BH/k_BT_f$ is the reduced magnetic field. Figure \ref{fig7} shows
 a comparison between the experimental measurements obtained in $\rm CuGa_2O_4$ and
Eq.\ref{mh}. There is a good agreement between theory and experiment for reduced magnetic
fields below $h=0.6$. For larger values of the reduced magnetic field $h$, however, there is
a significant discrepancy between the observed and calculated values of the 
magnetization, 
which can be attributed to the fact that the SK theory 
is based on the mean-field approximation and makes use of infinite 
long-range interactions between the spins.
On the contrary, the magnetic exchange interactions in $\rm CuGa_2O_4$ have short-range
character. Moreover,  the domain of validity of the mean-field theory 
for a spin-1/2 compound
is unclear, in particular close to the transition temperature ~$\rm T_f$
where  the critical fluctuations
become important. In the same spirit, the freezing
temperature $\rm T_f$ shows a much more pronounced magnetic-field 
dependence than predicted by mean-field theory
\begin{equation}
h^{2/3}=[4/(m+2)]^{1/3}(1-T_f(H)/T_f(0)),
\end{equation}
as evidenced in Fig.\ref{fig8}.
\section{$\mu SR$ results}
To get more insight into both the static and dynamic properties of the $\rm Cu^{2+}$
magnetic moments in $\rm CuGa_2O_4$, we have measured the muon-spin relaxation above and
below the freezing temperature $\rm T_f$ in this material.
Generally, the magnetic interactions probed by the implanted spin-polarized muon are 
detected by monitoring
the asymmetric emission of positrons arising from the weak decay of the muon. 
Recording the positron rate
$N(t)$ as a function of muon life-time yields
\begin{equation}
N(t)=N(0)\exp(-t/\tau)[1+AG_z(t)]~,
\label{rate}
\end{equation}
where $A$ is the initial muon asymmetry parameter. The product $AG_z(t)$ is 
often called the $\mu$SR signal.
In addition, the function $G_z(t)$ can be associated with the muon spin auto-correlation 
function, i.e. 
\begin{equation}
G_z(t)=\frac{\langle \mathbf{S}(t)\mathbf{S}(0) \rangle}{S^2(0)},
\end{equation}
where $\mathbf S$ is the spin of the muon.
%
Typical zero-field $\mu$SR signals measured in CuGa$_2$O$_4$ are shown in Fig. 
\ref{Fig9}. Above $T \simeq
3.8$~K (and at least up to 10~K), the data are best described by assuming for 
$G_z(t)$ the form
 \begin{equation}
  G_{z,para}(t)=G_{KT}(t)\cdot G_{es}(t)~,
  \label{gz}
 \end{equation}
with $G_{KT}$ representing the familiar Kubo-Toyabe (KT) expression \cite{Kubo}
\begin{equation}
  G_{KT}(t)=\frac{1}{3}+\frac{2}{3}(1-\Delta_{ns}^2t^2)
   \exp(-\frac{1}{2}\Delta_{ns}^2t^2)~,
 \end{equation}
and with $G_{es}$ given by 
\begin{equation}
  G_{es}(t)=\exp[-(\lambda t)^{\beta}]~.
 \end{equation}
The form of $G_{z,para}(t)$ points for the occurrence of two independent channels 
of depolarization 
acting on the muon spin. The first channel, giving rise to the KT function 
$G_{KT}$, originates 
from the nuclear dipole moments (Ga and Cu isotopes). The internal fields of 
this contribution 
are assumed to be Gaussian distributed in their values, randomly oriented and 
static within the 
$\mu$SR time window. The parameter $\Delta_{ns}^2/\gamma^2_{\mu}$ represents the 
second moment of 
this field distribution due to the nuclear spins along one Cartesian axis 
($\gamma_{\mu}$ = 
2$\pi\cdot$13.553879 kHz/G is the gyro-magnetic ratio of the muon). The second 
channel, 
described by the function $G_{es}(t)$, which will be discussed in details below, 
represents the 
contribution arising from the fluctuating electronic Cu spins.

At $T =10$K the muon depolarisation can be satisfactorily described by assuming 
$G_{es}(t) = 1$, i.e. 
$G_{z,para}(t) = G_{KT}(t)$ with $\Delta_{ns} = 0.16(1)$~MHz (see also Fig. 
\ref{Fig9}), 
indicating that the fluctuations of the electronic spins are still too fast to 
be observed in the 
$\mu$SR time window. However, upon cooling the sample below $T=10$~K and down to 
3.8~K, the 
fluctuation rate of the electronic spins decreases and the muon-spin depolarisation 
becomes gradually 
dominated by the $G_{es}(t)$ contribution.  Figure~\ref{Fig10} represents the 
temperature evolution of the 
depolarisation rate  $\lambda$. Whereas the exponent $\beta$ remains constant in 
this temperature 
interval (i.e. $\beta \simeq 0.78$), the depolarisation rate exhibits a marked 
critical-like divergence, 
which must be taken as a clear evidence of the approach to a magnetic phase transition 
as the temperature is 
decreased. This critical behavior at $\rm T\simeq 2.5K$ indeed corresponds to the 
temperature of the cusp in the magnetic susceptibility and to the specific
heat anomaly. It can therefore be associated to
the occurrence of the spin-glass phase (see Figs. \ref{fig1} and \ref{Fig6}) in $\rm CuGa_2O_4$. 
\\
For spin-glass systems, the stretched exponential form for the electronic-spin 
contribution of the 
muon-spin depolarisation function $G_z(t)$ has been shown \cite{campbell} to 
match the Kohlrausch-like 
stretched exponential for the local moments autocorrelation function itself, 
which in turn arises from a 
broad distribution of electronic-spin correlation times. In the particular case 
of moderately concentrated 
systems, the exponent $\beta$ reaches the value of $\frac{1}{3}$ at $T_f$. On 
the other hand for 
conventional magnetic systems, the dynamic muon-spin depolarisation function 
assumes an exponential form 
(i.e. $\beta = 1$), reflecting the unique spin-relaxation frequency of the 
localized moments. The situation 
observed here for CuGa$_2$O$_4$ appears somewhat intermediate with an exponent 
$\beta$ slightly, but
definitively, 
below unity ($\simeq \frac{3}{4}$). This behavior is tentatively ascribed to 
the high concentration 
 of local moments (Cu$^{2+}$ ions), randomly distributed in different 
sublattices, for which a 
somewhat narrow distribution of electronic-spin correlation times could be 
expected.
\\
In the temperature range between 3.8 and 10~K, the best fits with Eq. \ref{gz} 
provide 
a parameter $\Delta_{ns}$ for the KT function (i.e. essentially the width of the 
internal fields arising 
from the nuclear moments) which is practically constant, indicating that the 
nuclear moments remain static
at all temperatures. This is also confirmed by measurements performed in applied 
longitudinal fields (LF). 
If the nuclear moments are static within the $\mu$SR time window and if the 
applied field is sufficiently 
strong to quench the nuclear dipole field contribution (i.e. $G_{KT}(t) =1$), 
the muon-spin depolarisation 
should arise solely from the dynamical electronic-spin contribution and the 
depolarisation function will 
assume the form $G_z(t)=G_{es}(t)$.  This was indeed observed during LF 
measurements (see Fig. \ref{Fig11}) for 
which a magnetic field of 0.02~T was sufficient to quench the nuclear dipolar 
moments. The muon 
depolarisation function is then well reproduced with the stretched exponential 
function described before, 
with parameters compatible with the ones extracted from the zero-field data.

For temperatures below $T =3.8$~K, the muon depolarization increases 
significantly and assumes a Gaussian 
character at short times. For this temperature range, the best description of 
the data is obtained using 
the function 
 \begin{equation}
  G_z(t) = A_{para}G_{z,para}(t) + A_{magn}G_{DKT}(t)~,
  \label{gzmagn}
 \end{equation}
where $G_{z,para}(t)$ is defined above and $G_{DKT}(t)$ is the so-called 
dynamical Kubo-To\-yabe (DKT) 
function \cite{Kubo}, which reflects that the Gaussian internal field 
distribution due to the occurrence 
of static electronic-spins (second moment $\Delta_{es}^2/\gamma^2_{\mu}$) 
fluctuates at the rate
$\nu$. The first term of Eq. \ref{gzmagn} is only present in the temperature 
interval between 3.8 and 2.5~K,
i.e. in a region where paramagnetic domains appear to coexist with domains 
exhibiting static, albeit
disordered, magnetic moments. Figure~\ref{Fig12} shows the temperature evolution 
of the amplitude $A_{magn}$
which mirrors the volume of the magnetic domains. Therefore, it appears that in 
CuGa$_2$O$_4$ the transition
to a spin-glass state begins around $T \simeq 3.8$~K to form local clusters of 
frozen electronic spins which
grow when the temperature is lowered and finally percolate at the same temperature 
where the specific
anomaly is observed (i.e. $T \simeq 2.5$~K) and which can be therefore 
associated to $T_f$.\\
With the exception of some limiting cases, the DKT function cannot be expressed 
analytically and depends
directly on the parameters $\nu$ and $\Delta_{es}$. Figure \ref{Fig13} shows the
temperature dependence of the parameter $\Delta_{es}$ which exhibits a clear 
increase below $\simeq 3.5$~K and
can be
associated to the temperature dependence of the static part of the electronic 
magnetic moments. The
fluctuation rate $\nu$ was found to be constant below $T_f$ ($\nu \simeq 3.7$ 
MHz). It is worthwhile to note
that the DKT function, which appears to describe perfectly the data for $T\ll 
T_f$, assumes a single
fluctuation rate $\nu$ for the internal fields sensed by the muon spin. This has 
to be connected to our simple
picture that the slightly reduced value in the paramagnetic phase of the 
exponent $\beta$ compared to unity
must be related to a rather narrow distribution of electronic-spin correlation 
times. 
Interestingly, the DKT function describes the data more satisfactorily than the 
model of
``coexisting static and dynamical fields'' developed by Uemura {\it et al.} 
\cite{Uemura}, and based on the
theory of Edwards and Anderson \cite{EA}, where each local moment is taken as a 
vector sum of a 
static component and a dynamical component randomly fluctuating (see Fig.\ref{Fig14}).
 This indicates that in CuGa$_2$O$_4$, by decreasing the temperature, 
an increasing part of each Cu$^{2+}$ moment become quasi-static
(characterized by a slow fluctuation rate $\nu$), whereas the remaining part 
does not affect the muon
polarization due to fast fluctuations that are not accessible within the $\mu$SR time 
window.
\section{Monte-Carlo simulations}
The experimental observations presented in the preceding sections all indicate
that the $\rm Cu^{2+}$ magnetic moments in $\rm CuGa_2O_4$ undergo a phase
transition to a spin-glass state at $\rm T_f$=2.5 K. To understand the nature of 
this magnetic state, Monte-Carlo simulations were performed using a model of Heisenberg
spins with competing exchange interactions including random anisotropies. These arise
as the result of Jahn-Teller distortions of the octahedrons and 
tetrahedrons surrounding the $\rm Cu^{2+}$ positions \cite{kugel}. The local distortions occur
randomly along one of the three equivalent $\rm C_4$ cubic axes. Consequently, 
the exchange interactions between nearest-neighbors spins located on tetrahedral
(A-sites) and octahedral (B-sites) positions have tetragonal anisotropy.
However, the direction of the tetragonal axis is random in a crystal with cubic symmetry.
For the model calculations,  we considered exchange 
interactions between nearest-neighbors $\rm Cu^{2+}(A)-Cu^{2+}(B)$ and second-nearest-
neighbors $\rm Cu^{2+}(B)-Cu^{2+}(B)$ magnetic ions. We took into account the fact
that in the spinel lattice the $\rm Cu^{2+}$ ions are randomly distributed
 between the A and B sites with occupation probabilities of 25\% and 75\%, respectively.
Consequently, the model Hamiltonian for this spin system is given by
\begin{eqnarray}
\mathbf{H}& =& -\sum_{\alpha=x,y,z}\sum_{i,j}J_{ij}^{\alpha\alpha}S_i^\alpha S_j^\alpha
P_i^tP_j^0
\nonumber\\
& & -\sum_{i,j}K_{ij}S_iS_jP_i^tP_j^0-\sum_iHS_i^z(P_i^tP_j^0),
\end{eqnarray}
where the $J_{ij}^{\alpha\alpha}${}'s represent the exchange integrals between the
nearest-neighbors $\rm Cu^{2+}$ ions located in the A and B sites; $K_{ij}$ is the
exchange parameter between nearest-neighbors $\rm Cu^{2+}$ ions on the octahedral
sublattice and H the external magnetic field. The components of the 
exchange interactions $J_{ij}^{\alpha \alpha}$ are 
distributed randomly with the same probability, namely
\begin{eqnarray}
P(J_{ij}^{xx},J_{ij}^{yy},J_{ij}^{zz})= & {} &
\nonumber\\
 1/3\delta(J_{ij}^{xx}-J_0-\Delta J)
\delta(J_{ij}^{yy}-J_0)\delta(J_{ij}^{zz}-J_0)&+&
\nonumber\\
 1/3\delta(J_{ij}^{xx}-J_0) 
\delta(J_{ij}^{yy}-J_0-\Delta J)\delta(J_{ij}^{zz}-J_0)&+&
\nonumber\\
 1/3\delta(J_{ij}^{xx}-J_0)
\delta(J_{ij}^{yy}-J_0)\delta(J_{ij}^{zz}-J_0-\Delta J),& &
\end{eqnarray}
where $\delta (x)$ is the $\delta$-function. The random numbers $P_i^t$ and $P_j^0$ 
determine the distribution \textbf{P} of the $\rm Cu^{2+}$ ions among the tetrahedral
 and octahedral sites in the spinel lattice, respectively, so that
\begin{equation}
{\mathbf{P}}(P_i^{t,0})=\nu^{t,0}\delta(P_i^{t,0}-1)+(1-\nu^{t,0})\delta (P_i^{t,0})
\end{equation}
with $\nu^t=0.25$ and $\nu^0=0.75$. The Monte-Carlo simulations were carried out using
periodic boundary conditions for a lattice consisting of $24\times 24\times 24$ sites
and over 30000-60000 MK steps per spins. We calculated the
magnetization of the spin lattice, the magnetic susceptibility, and the spin-spin correlation 
function $\rm \langle S(0)S(R) \rangle$. We simulated the temperature
dependence of the EA-order parameter q(T) for
the A- and B-spins, respectively, defined  as
\begin{equation}
q^{\alpha\beta}=(1/N)\sum_{i=1}\langle S_i^\alpha\rangle ^2,{ } \alpha=x,y,z,{ } \beta=t,0. 
\end{equation}
The exchange parameters $\Delta J$, $K$ and $J_0$ were obtained
 by fitting the MK results
 to the experimental freezing temperature $\rm T_f$, the paramagnetic 
N\'eel temperature $\Theta$ and the magnetic field dependence of the magnetization M(H).
Fig. \ref{fig15} shows the temperature dependence of the magnetic susceptibility
calculated by the Monte-Carlo method for two values of magnetic fields, H=0 Oe 
and H=$10^4$ Oe, respectively. In agreement with the experimental results, the 
calculated magnetic susceptibility exhibits a cusp at $\rm T_f\simeq 2.5K$ which
is suppressed when a magnetic field is applied. As shown in Fig.\ref{fig16} the EA-order
parameters for both the A- and B-sublattice sharply increase below T=$\rm T_f$.
Moreover, the spin-spin correlation function $\rm \langle S(0)S(L/2) \rangle$ 
(L=Monte-Carlo sample size) reveals the absence of any long-range magnetic ordering
in the spin system. The Monte-Carlo results show that for the concentrations of
$\rm Cu^{2+}$ spins of relevance for $\rm CuGa_2O_4$, the crystal is in a 
superparamagnetic state when K=0. Introducing random anisotropy for
the exchange interaction J results in a spin cluster blocking  at 
the freezing temperature $\rm T_f$. In that respect, we note that the freezing temperature is 
proportional to $\Delta$J. To reproduce the magnetization data in a satisfactory
way, we found necessary to give a non-zero value to the 
antiferromagnetic exchange interaction K. From a least-square refinement of 
the field dependence of the magnetization we obtained the parameter values
$\Delta J/J_0$=0.1, $K/J_0$=0.5. The exchange parameter $J_0$, as determined from the 
freezing temperature $T_f$, from the paramagnetic susceptibility in the temperature range
$\rm 90K\le T\le 160K$ and from the magnetization curve, amounts to 
-12K, -12.5K and -13 K, respectively. Therefore, the mean values of the model parameters
are $J$=-12.5K, $\Delta J$=-1.3K, and $K$=-6.2K.  

\section{Conclusion}
We have presented magnetization, magnetic susceptibility, specific heat 
and \musr\ measurements in $\rm CuGa_2O_4$. The data are consistent with a spin-glass
transition of the copper sublattice below $\rm T_f$ in this material. In particular, 
we observe a cusp in the temperature dependence of the magnetic susceptibility
at $\rm T_f\simeq 2.5K$ which is suppressed when a magnetic field is applied.
A pronounced hysteresis is observed in the temperature dependence of the magnetic
susceptibility for zero-field and field-cooled samples. The muon-spin
relaxation measurements have shown that above the 
freezing temperature, the
asymmetry function  is described by the stretched exponential 
typical of disordered
systems. However, the value of the exponent $\beta$ points to a narrow 
distribution of correlation times of
the local moments. The temperature dependence of the magnetic volume fraction 
indicates that in $\rm CuGa_2O_4$ the
transition to a spin-glass state begins around T$\simeq$3.8K to form locally 
clusters of frozen spins 
which grow when the temperature is lowered and finally percolate around $T_f 
\simeq 2.5$~K. By means of Monte-Carlo simulations we were able to reproduce the
main features of
the magnetic susceptibility and of the magnetization curve measured in $\rm CuGa_2O_4$.
The results of Monte-Carlo simulations show that the Jahn-Teller effect plays an essential role 
in forming the magnetic ground-state as it introduces random anisotropy in the exchange 
interactions between the copper ions, which in turn is responsible for the formation 
of a spin-glass state in $\rm CuGa_2O_4$.
Using a realistic spin model which takes into account the effective distribution of the 
$\rm Cu^{2+}$ ions in the spinel structure, 
reliable exchange parameters could be obtained for $\rm CuGa_2O_4$.
\section{Acknowlegments}
The work is partially supported by INTAS-97-0177 grant and by CICYT Project 99/142.

*On leave of absence in the Kamerlingh Onnes Laboratorium, Leiden
University, The Netherlands.
\begin{figure}
\caption{(a) Real  and (b) imaginary  parts of the magnetic susceptibility in a single 
crystal of  $\rm CuGa_2O_4$ as measured with a SQUID magnetometer at a frequency of 19 Hz and in 
an applied magnetic field of 4.5 Oe.}
\label{fig1}
\end{figure}
\begin{figure}
\caption{Experimental value of the magnetic moment in $\rm CuGa_2O_4$ for a sample
 cooled in an applied magnetic field of H=100 Oe (FC curve) and in zero-field (ZFC curve).}
\label{fig2}
\end{figure}
\begin{figure}
\caption{Temperature dependence of the magnetic susceptibility for different magnetic fields.}
\label{fig3}
\end{figure}
\begin{figure}
\caption{Curve 1 shows the magnetic field dependence of the magnetic moment in $\rm CuGa_2O_4$
at the temperature of T=1.8 K. The magnetic field is applied along the [001] crystal axis.
In curve 2, we show the result of the Monte-Carlo simulation for the same field dependence.}
\label{fig4}
\end{figure}
\begin{figure}
\caption{Magnetic susceptibility measured in $\rm CuGa_2O_4$ for different frequencies.} 
\label{fig5}
\end{figure}
\begin{figure}
\caption{Specific heat of $\rm CuGa_2O_4$.} 
\label{Fig6}
\end{figure}
\begin{figure}
\caption{Temperature dependence of M/H in $\rm CuGa_2O_4$ at T=1.8 K.} 
\label{fig7}
\end{figure}
\begin{figure}
\caption{Experimental and theoretical dependences of the freezing temperature $\rm T_f$
as a function of the reduced magnetic field h, as explained in the text.} 
\label{fig8}
\end{figure}
\begin{figure}
\caption{ Experimental zero-field $\mu$SR signal measured in $\rm CuGa_2O_4$ 
at $\rm T = 10$~K, $\rm T = 4.5$~K and $\rm T=650$~mK. 
The lines represent fits as explained in the text.}
\label{Fig9}
\end{figure}
\begin{figure}
\caption{Temperature dependence of the depolarisation rate $\lambda$ above 
3.8~K.}
\label{Fig10}
\end{figure}
\begin{figure}
\caption{$\mu SR$ spectra measured in $\rm CuGa_2O_4$ at $\rm T = 5$~K and showing the 
field dependence of the asymmetry function. Whereas in zero-field the 
depolarization function is best
described according to Eq. \ref{gz}, the data with applied field are only 
fitted by the stretched
exponential term reflecting the depolarisation arising from the fluctuating 
electronic spins (see text).}
\label{Fig11}
\end{figure}
\begin{figure}
\caption{Temperature dependence of the parameter $A_{magn}$ corresponding to the 
magnetic volume fraction.}
\label{Fig12}
\end{figure}
\begin{figure}
\caption{Temperature dependence of the $\Delta$-parameter of the DKT function. 
This parameter mirrors the
width of the quasi-static field distribution below $\rm T_f$ and therefore the value 
of the quasi-static
Cu$^{2+}$ moment.}
\label{Fig13}
\end{figure}
\begin{figure}
\caption{Fit to a $\mu$SR spectrum using the Uemura's function 
\cite{Uemura} (see also text).
For sake of clarity the error bars are omitted. The shortcomings of 
the fit are evident.}
\label{Fig14}
\end{figure}
\begin{figure}
\caption{Temperature dependence of the normalized magnetic susceptibility
for values of magnetic fields H=0 Oe and H=$\rm 10^4$ Oe (curve 1 and 2, respectively) 
with exchange constants J=-12 K, K=-6 K and $\Delta$J=-1.2 K.}
\label{fig15}
\end{figure}
\begin{figure}
\caption{Temperature dependence of the Edwards-Anderson parameter q(T) for
the A- and B-sublattice as simulated with Monte-Carlo. See text for details.}
\label{fig16}
\end{figure}
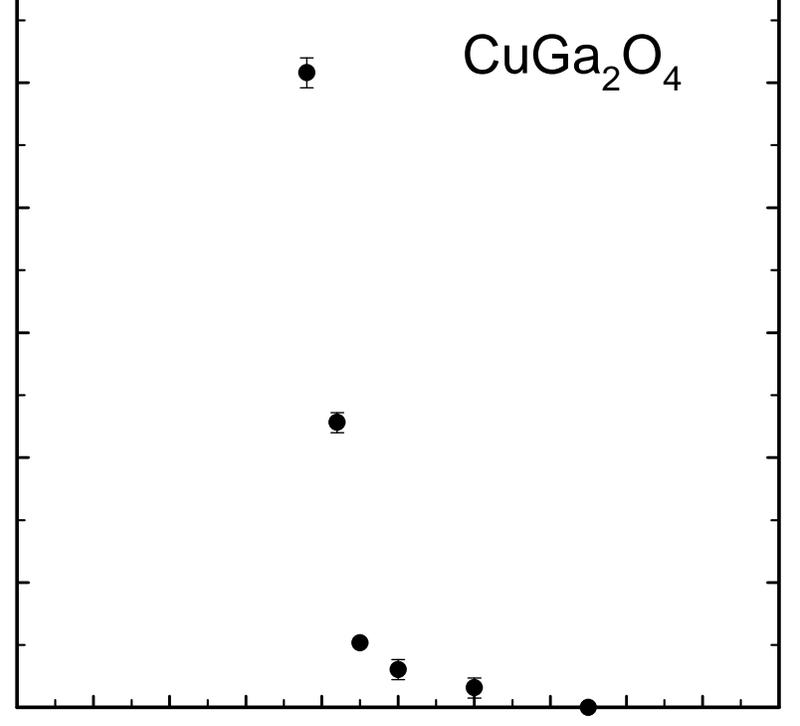
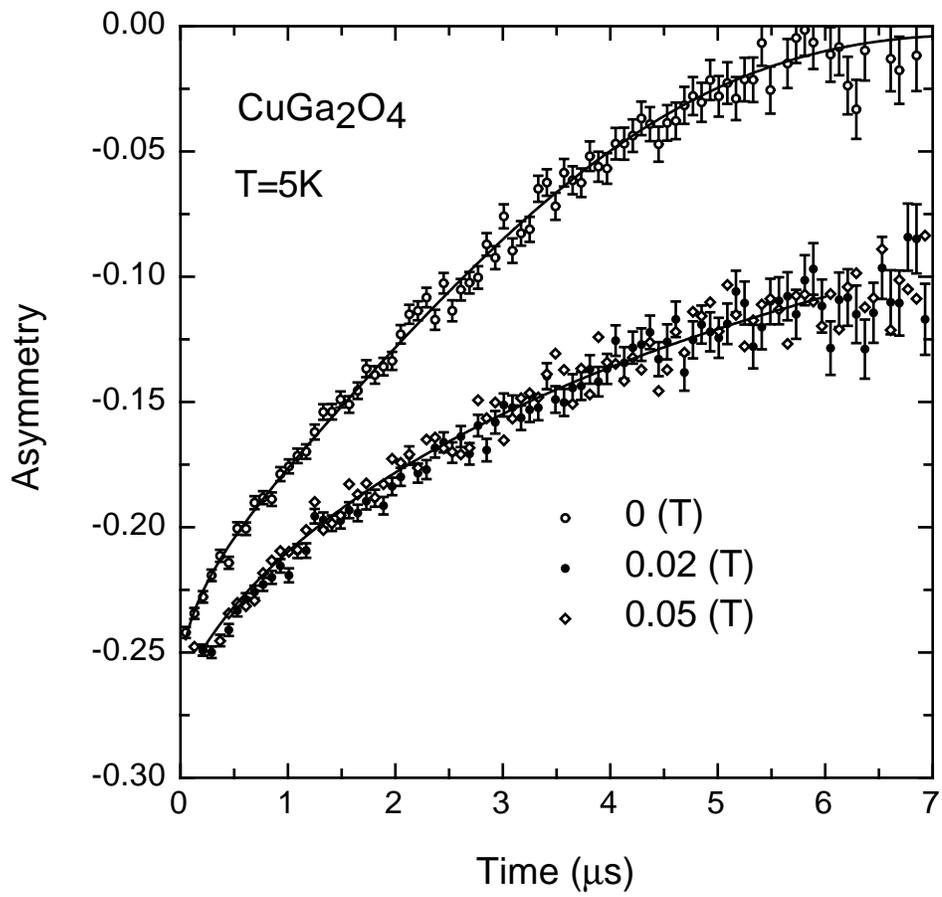
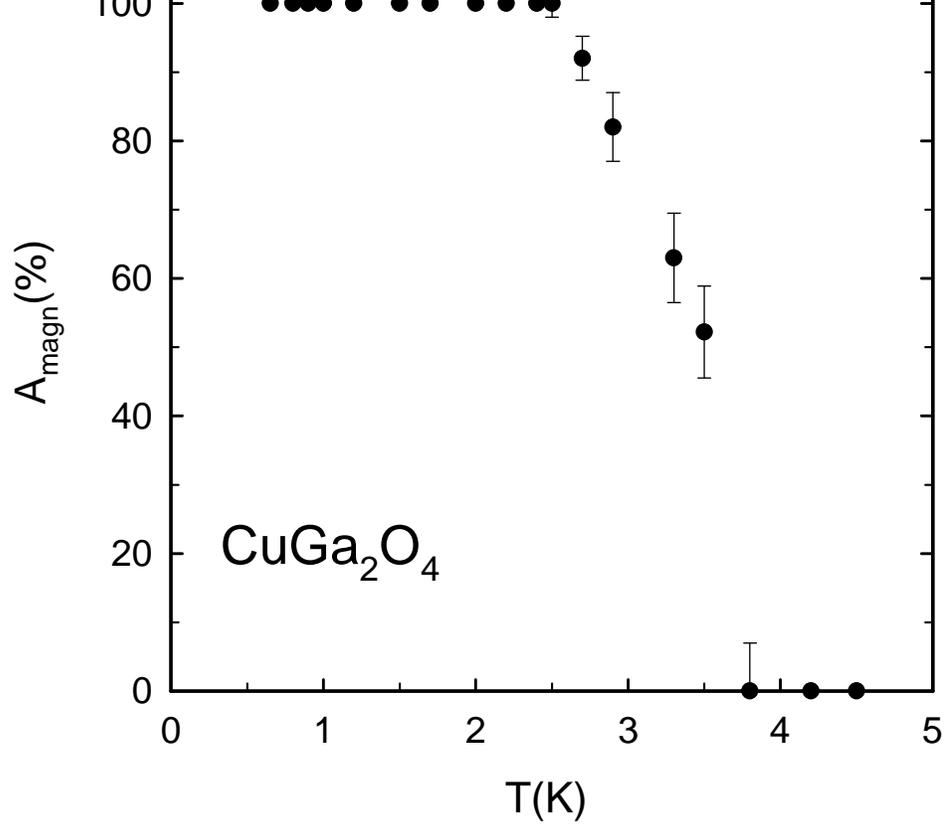
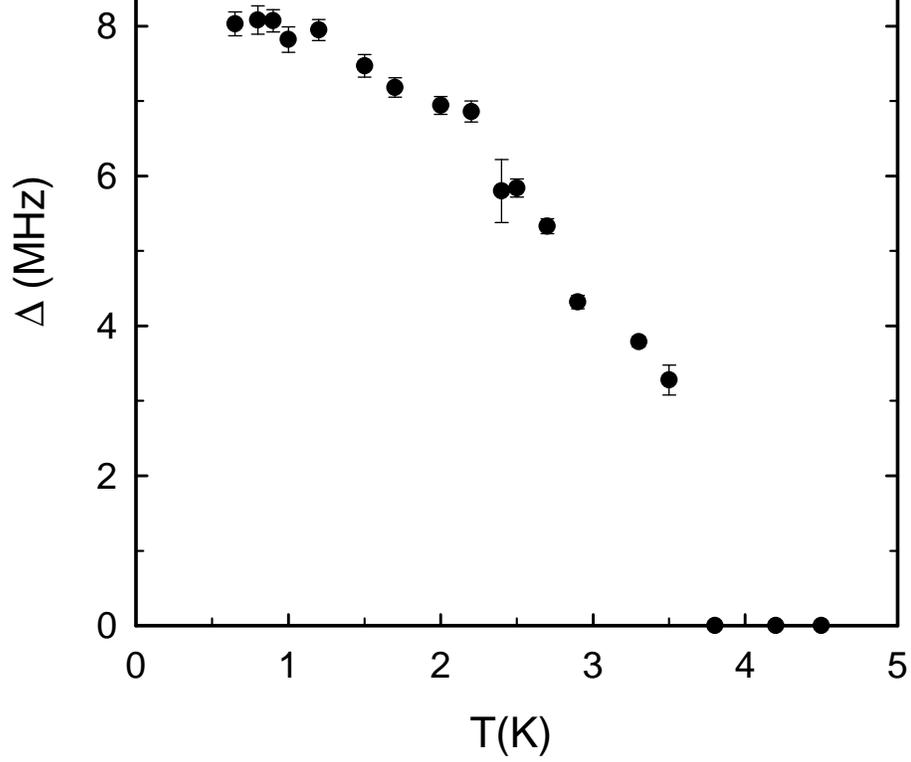
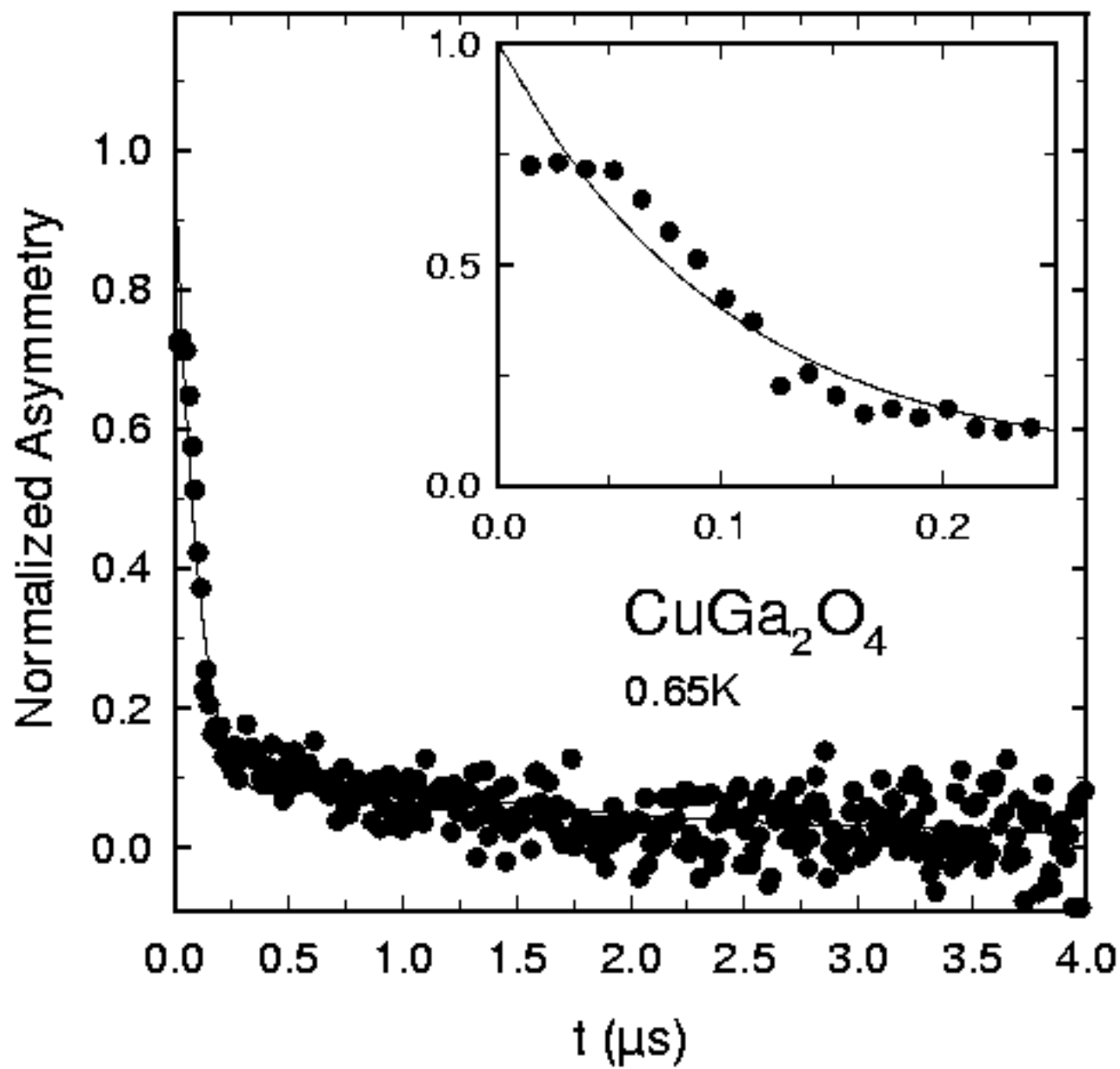
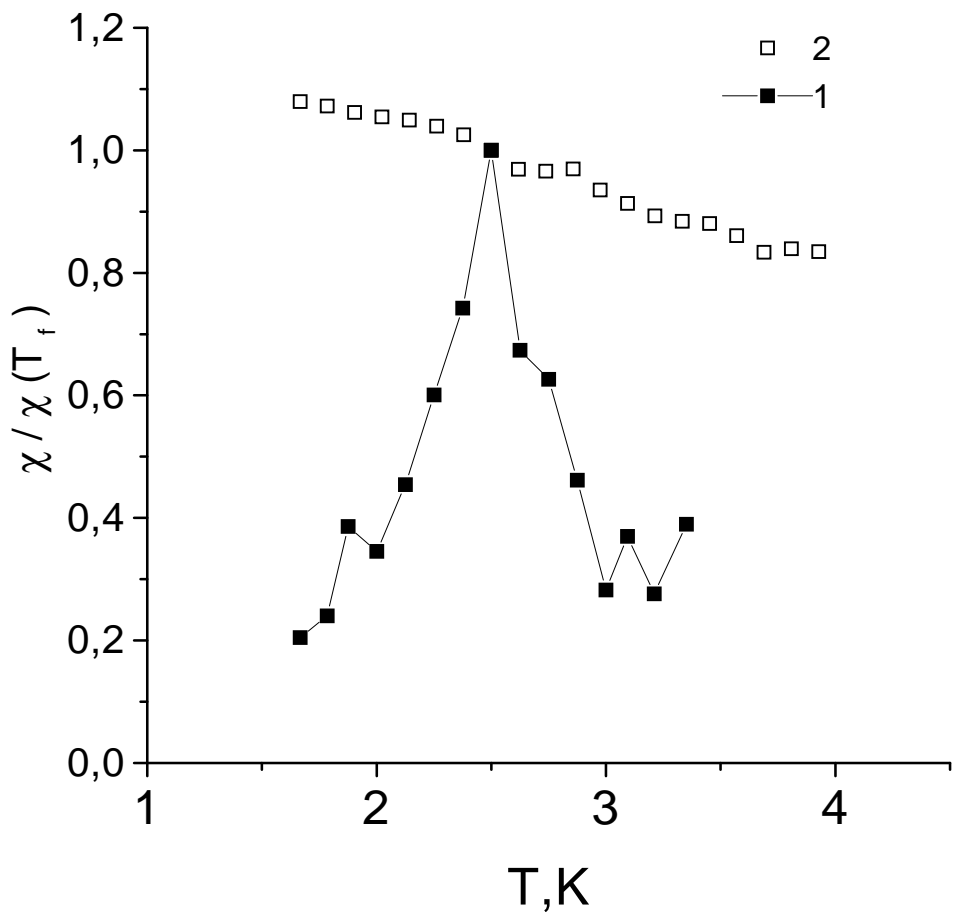
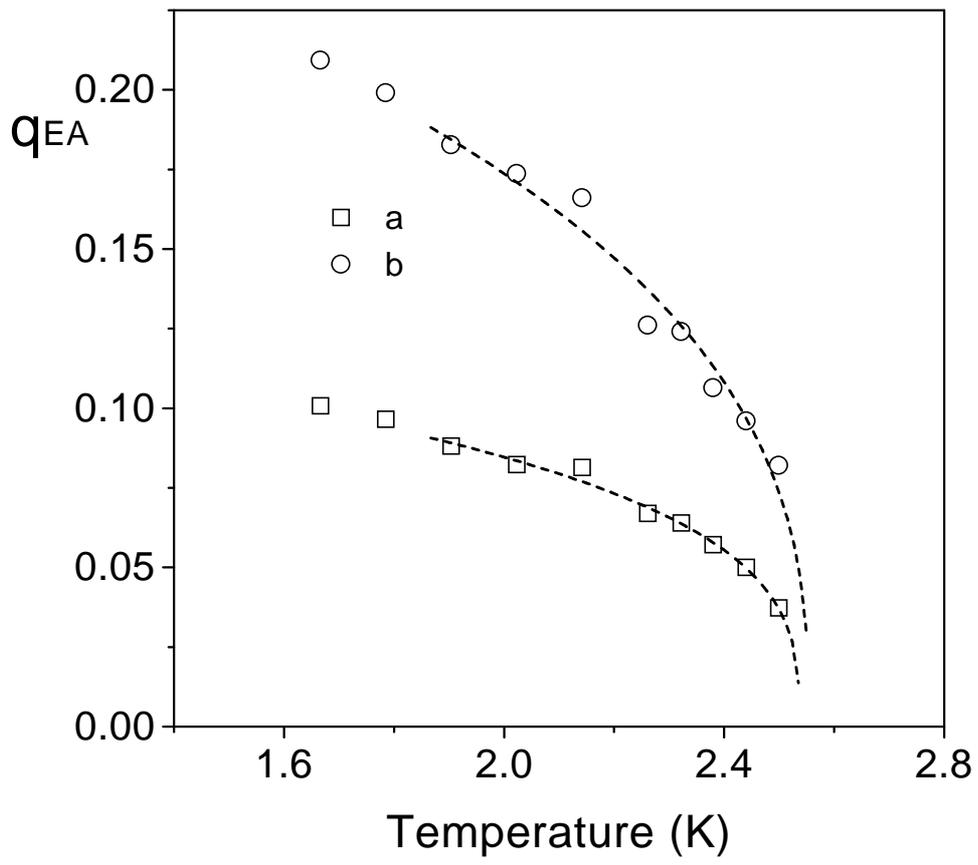
\end{document}